\documentclass[aps,pre,preprint,onecolumn,superscriptaddress]{revtex4-2}

\usepackage{url} %this package should fix any errors with URLs in refs.
\usepackage{graphicx} 
\usepackage{subfig}
\usepackage{dblfloatfix} 
\usepackage{siunitx}

\usepackage{amsmath}
\usepackage{amssymb,bm,exscale}

\begin{document}

\title{Flow-rate controls on trapped hydrogen redistribution in a heterogeneous sandstone: a synchrotron X-ray micro-CT study}

\author{Catherine Spurin}
\affiliation{Department of Energy Science \& Engineering, Stanford University}
\email{cspurin@stanford.edu}

\author{Sojwal Manoorkar}
\affiliation{Center for X-ray Tomography \& Department of Geology, Ghent University, Belgium}

\author{Sharon Ellman}
\affiliation{Center for X-ray Tomography \& Department of Geology, Ghent University, Belgium}

\author{Ian B. Butler}
\affiliation{School of Geosciences, University of Edinburgh, James Hutton Road, The King’s Buildings, Edinburgh EH9 3FE, United Kingdom}
\affiliation{University of Manchester, Diamond Light Source, Harwell Campus, Didcot OX11 0DE, UK}

\author{Aliakbar Hassanpouryouzband}

\author{Damien Freitas}
\affiliation{School of Geosciences, University of Edinburgh, James Hutton Road, The King’s Buildings, Edinburgh EH9 3FE, United Kingdom}
\affiliation{University of Manchester, Diamond Light Source, Harwell Campus, Didcot OX11 0DE, UK}
\author{Fernando Alvarez-Borges}
\affiliation{\textmu-VIS X-ray Imaging Centre, University of Southampton, Southampton SO17 1BJ, UK}
\author{Robert Atwood}
\affiliation{Diamond Light Source, Harwell Campus, Didcot OX11 ODE, UK}
\author{Katriona Edlmann}
\author{Eike Thaysen}
\affiliation{Department of Geosciences, Institute of Environmental Assessment and Water Research (IDAEA), Spanish Council for Scientific Research (CSIC), Jordi Girona 18–26, 08034 Barcelona, Spain}

\begin{abstract}

Subsurface hydrogen storage in porous geological formations is a promising strategy
for balancing temporal mismatches between renewable energy generation and
demand. The viability of large-scale hydrogen storage depends on predicting hydrogen
distribution, trapping, and recovery during cyclic injection and withdrawal operations.
However, these multiphase flow processes are influenced by rock heterogeneity,
capillary hysteresis and trapping at the pore-scale. In this work, fast in-situ synchrotron
X-ray tomography was used to directly visualise hydrogen mobilisation during brine
imbibition in a heterogeneous sandstone. By systematically varying the brine injection
rate, the influence of flow conditions on hydrogen recovery, and the evolution of
trapped hydrogen were quantified. Significant mobilisation of disconnected hydrogen
clusters was observed during imbibition, even at low capillary numbers ($< 10^{-6}$).
Rock heterogeneity influenced trapping behaviour, particularly during the early stages of imbibition. Our results highlight the importance of incorporating pore-scale
heterogeneity and transient trapping mechanisms into hydrogen storage predictions.

\end{abstract}

\maketitle

%%%%% MAIN TEXT 
\section{Introduction}
Many renewable energy sources, such as solar and wind, are inherently intermittent, with their output governed by fluctuating factors such as solar irradiance and wind intensity \cite{heide2010seasonal, engeland2017space}. This variability presents a significant challenge for balancing energy supply and demand \cite{schaber2012parametric}. Hydrogen offers a promising low-carbon energy carrier that can help address this issue. In particular, “green hydrogen,” produced via electrolysis during periods of excess renewable generation, enables surplus energy to be stored and later utilized during times of shortage, thereby enhancing the flexibility and resilience of energy systems \cite{lord2014geologic, ali2025recent, international2022global, raza2022holistic}. Batteries, storage tanks, or salt caverns can provide temporary  storage, but their capacity is limited \cite{andersson2019large, ozarslan2012large, bai2014overview}. Subsurface storage in porous geological reservoirs offers the potential to store and recover large volumes of hydrogen on demand \cite{krevor2023subsurface, hematpur2023review}. Understanding hydrogen migration and trapping mechanisms in the subsurface is required to accurately predict storage capacity and recovery efficiency.

The spatial distribution of hydrogen during injection, and its subsequent recovery is strongly influenced by reservoir heterogeneity \cite{jangda2024subsurface, pfeiffer2017hydrogen, mosallanezhad2026coupled}. While numerous studies have explored the impact of heterogeneity on multiphase flow for other fluid pairs such as nitrogen-water or CO$_2$-water \cite{spurin2019mechanisms, darraj2026impact, soltanmohammadi2024investigation, zahasky2020pore, wang2023pore}, hydrogen has been shown to exhibit distinct displacement, trapping, and recovery behaviour \cite{zhang2026pore, heinemann2021enabling, zhang2023pore, thaysen2023pore}. Unlike geological CO$_2$ storage, where the primary objective is permanent containment, hydrogen storage requires repeated injection and recovery cycles to balance variations in energy supply and demand. These operational cycles occur on much shorter timescales than those associated with conventional subsurface storage or hydrocarbon production, making the remobilisation and recovery of trapped hydrogen a key consideration. The amount of recoverable hydrogen has been observed to evolve over successive withdrawal cycles due to capillary pressure hysteresis and changes in fluid connectivity, resulting in progressive changes in recovery efficiency \cite{jangda2024subsurface, zhang2024effect, gao2023pore,spurin2025cwtcyclic}. Previous pore-scale investigations have demonstrated that hydrogen recovery generally increases with increasing brine flow rate, reflecting the competition between viscous mobilization and capillary trapping forces \cite{thaysen2023pore}. Residually trapped hydrogen commonly occupies ganglia spanning multiple pores, emphasizing the critical role of capillary trapping and pore-scale connectivity in governing storage performance \cite{thaysen2023pore}. Furthermore, hydrogen recovery dynamics span a broad range of timescales, including rapid displacement events and slower pore-scale rearrangements driven by processes such as Ostwald ripening \cite{zhang2023pore, salehpour2026ostwald, goodarzi2024trapping}.

Understanding the mechanisms governing hydrogen trapping is essential for predicting the long-term performance of subsurface hydrogen storage and recovery operations. Previous work has examined pore-scale hydrogen displacement using laboratory-based micro-CT imaging \cite{jangda2024subsurface, zhang2023pore, thaysen2023pore, jha2021pore, dokhon2024pressure, song2025visualized, al2024snap, goodarzi2024trapping}. Fast synchrotron tomography has been used to investigate hydrogen injection and displacement \cite{thaysen2025time}; however, the temporal evolution of hydrogen mobilisation during brine invasion, and the underlying pore-scale mechanisms governing its displacement and remobilisation, have not, to our knowledge, been systematically investigated. Furthermore, observations derived from other gas-water systems cannot be assumed to apply directly to hydrogen. Unlike nitrogen and CO$_2$, hydrogen possesses a lower viscosity, higher diffusivity, and lower solubility in water; properties that are expected to significantly influence displacement dynamics, phase connectivity, and residual trapping behaviour \cite{raza2022holistic, lysyy2022pore, zhang2023pore, manoorkar2025pore, zhang2026pore}. These distinctive characteristics highlight the need for dedicated experimental studies using hydrogen to develop a mechanistic understanding of its storage and recovery behaviour in heterogeneous porous media.

In this work, fast in-situ synchrotron X-ray tomography was used to investigate the temporal evolution of hydrogen recovery in a heterogeneous sandstone as a function of brine flow rate. Unsteady-state experiments were performed at a pore fluid pressure of 5~MPa and a confining pressure of 10~MPa, representative of shallow subsurface hydrogen storage conditions \cite{yekta2018determination, okoroafor2022toward}. The objectives were to: (1) identify the onset of hydrogen mobilisation under bulk capillary-dominated flow conditions; (2) characterise the temporal evolution of hydrogen recovery and phase connectivity as a function of flow rate; (3) quantify the role of heterogeneity in controlling the mobilisation of residually trapped hydrogen; and (4) establish the relationship between hydrogen recovery and brine invasion rate.

\section{Materials and Methods}

\subsection{Core-flooding experiments}
The experiments were conducted in a cylindrical Clashach sandstone core, 4.7~mm in diameter and 54~mm in length. Clashach sandstone is a Permian aeolian sandstone from Scotland with a porosity ranging from 12 to 18\% \cite{thaysen2025time, jangda2024subsurface, ngwenya2003permeability, thaysen2023pore, ngwenya2000experimental}. The Clashach sandstone was selected for its relatively low mineralogical variability and comparatively large pore spaces, which facilitated the visualisation. The sample and experimental apparatus were identical to those described by \cite{thaysen2025time}. Briefly, the core was mounted in an X-ray transparent core holder equipped with two inlet ports and a single outlet port. Two syringe pumps independently injected H$_2$ and brine into the sample under constant-flow conditions, while a third pump maintained a constant pore pressure of 5~MPa at the outlet. A fourth pump applied a confining pressure of 10~MPa, corresponding to 5~MPa above the pore pressure, to prevent fluid bypass around the core.

Prior to the experiments, the sample was fully saturated with brine consisting of deionized water doped with 2~M KI to enhance X-ray phase contrast. The KI was added solely to improve phase discrimination during image segmentation; consequently, although the experiments capture representative pore-scale displacement processes, the elevated brine salinity and density should be considered when extrapolating the results directly to reservoir conditions. After saturation, the system was pressurized to a pore pressure of 5~MPa with a confining pressure of 10~MPa. All experiments were conducted at ambient temperature (293~K). Hydrogen was initially injected at a flow rate of \SI{20}{\micro\litre\per\minute}. Once hydrogen reached the imaged region, injection continued for an additional five pore volumes (PV) to establish the initial gas distribution. Hydrogen injection was then stopped, and brine was injected sequentially at flow rates of 5, 10, and \SI{20}{\micro\litre\per\minute} to investigate hydrogen remobilisation during imbibition. Each flow-rate condition was maintained for 34 minutes, such that the total injected brine volume increased with increasing flow rate. The lowest flow-rate condition still resulted in the injection of at least one pore volume, ensuring sufficient fluid displacement for assessing hydrogen remobilisation. The cumulative injected brine volumes corresponded to 1.33, 2.67, and 5.33 pore volumes (PV), respectively.

\begin{table}[!h]
\centering
\setlength{\tabcolsep}{15pt} % increase column spacing
\caption{Capillary numbers for the three brine injection rates.}
\label{ca-number}
\begin{tabular}{cc}
\hline
\textbf{Brine injection rate} & \textbf{Capillary number} \\
\textbf{(\si{\micro\litre\per\minute})} & \\
\hline
5  & $7.08 \times 10^{-8}$ \\
10 & $1.42 \times 10^{-7}$ \\
20 & $2.83 \times 10^{-7}$ \\
\hline
\end{tabular}
\end{table}

The capillary numbers reported in Table \ref{ca-number} indicate that all experiments were conducted in the capillary-dominated flow regime. The capillary number was calculated as:

\begin{equation}
Ca = \frac{q}{\sigma \lambda},
\label{Ca_number_eq}
\end{equation}

where $q$ is the Darcy flux (the total volumetric injection rate per unit cross-sectional area), $\sigma$ is the interfacial tension between H$_2$ and water (taken as \SI{72.6e-3}{\newton\per\metre} from \cite{chow2018interfacial}), and $\lambda$ is the total fluid mobility, defined as:

\begin{equation}
\lambda = \frac{f_w}{\mu_w} + \frac{1-f_w}{\mu_g},
\end{equation}

where $\mu_w$ is the viscosity of the water (taken as 1.07 $\times$ $10^{-3}$ Pa$\cdot$s from \cite{iglauer2012comparison}), $\mu_{g}$ is the viscosity of the hydrogen (taken as 9.01 $\times$ $10^{-6}$ Pa$\cdot$s from \cite{yusibani2011capillary}), and $f_w$ is the fractional flow of water \cite{Spurin2019phase}. Since all experiments were performed with pure brine injection ($f_w$ = 1), the mobility reduces to $1/\mu_w$, and the hydrogen viscosity does not contribute to the capillary number calculation.

Fluid distributions were imaged using synchrotron X-ray tomography at the Diamond Light Source (I12-JEEP beamline). Before hydrogen injection, a tomographic scan of the fully brine-saturated core was acquired to provide the reference image for image registration and phase segmentation. Following hydrogen injection, time-resolved tomographic images were acquired throughout the subsequent brine injection stages to capture fluid distributions and their evolution during imbibition. Multiple scans were collected at each flow rate, with scans equally spaced in time to investigate the dynamics of imbibition and the temporal evolution of hydrogen connectivity. Each tomographic scan consisted of 900 projections acquired with an exposure time of 0.017~s per projection using a 65~keV monochromatic X-ray beam, yielding a reconstructed voxel size of \SI{3.1}{\micro\meter}. To achieve near-continuous imaging, successive scans were collected using alternating clockwise and anticlockwise 180$^\circ$ half-rotations. Each tomographic scan took approximately \SI{15}{\second} to acquire, with \SI{5}{\second} between successive scans.

\subsection{Image analysis}

\begin{figure}[!htb]
    \centering
    \includegraphics[width=0.8\textwidth]{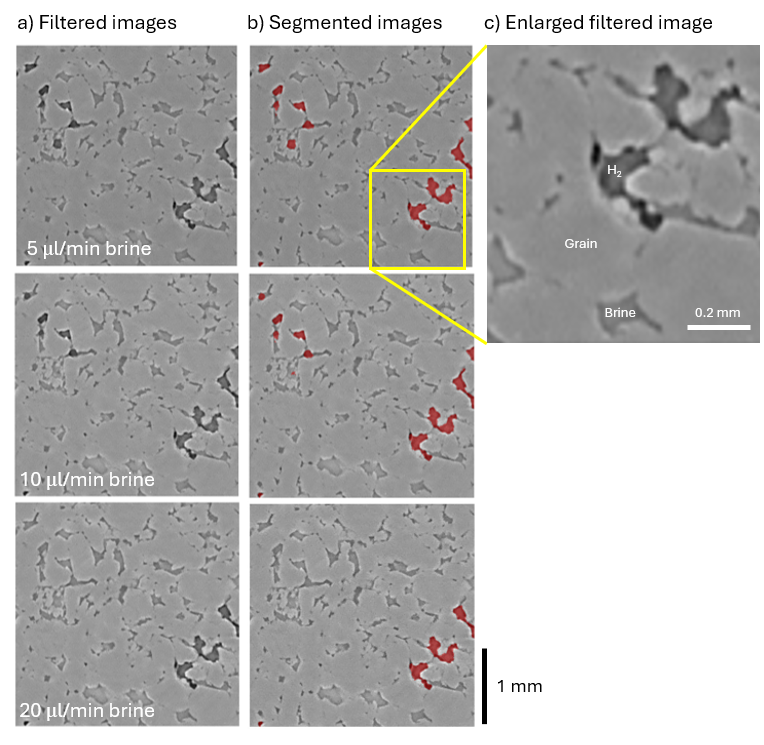}
    \caption{a) Non-local means filtered micro-CT images acquired during the progressive displacement of hydrogen from the rock, where hydrogen is the darkest phase, b) corresponding hydrogen segmentation (red) overlaid on the filtered images, and c) enlarged view of a filtered image highlighting the hydrogen, brine, and rock grain phases. Flow is into the page.}
    \label{fig:QC_segmentation}
\end{figure}

Image segmentation used the workflow described in \cite{spurin2024python}. Firstly, all images were filtered with a non-local means filter \cite{buades2011non}. Then all flow scans (scans containing both brine and H$_2$) were registered to the brine-saturated scan. Then H$_2$ segmentation was achieved through image subtraction, whereby the brine-saturated scan taken was subtracted from scans containing both brine and H$_2$, followed by threshold-based segmentation of the resulting difference image. Small ganglia (less than 500 voxels; which equates roughly to an object 8$\times$8$\times$8 voxels in size) were removed from the analysis to reduce dependence on image noise. A representative subvolume of \SI{2170}{\micro\meter} $\times$ \SI{2170}{\micro\meter} $\times$ \SI{2201}{\micro\meter} was analysed. Representative segmented hydrogen distributions for each brine injection rate are shown in Figure \ref{fig:QC_segmentation}. The selected examples illustrate the progressive displacement of H$_2$ from the upper-left region and the central-right portion of the sample during brine injection. The segmentation resolves both the macroscopic evolution of the hydrogen phase and localized pore-scale meniscus retraction events, demonstrating its ability to capture displacement processes across multiple length scales.

\begin{figure}[!htb]
    \centering
    \includegraphics[width=0.8\textwidth]{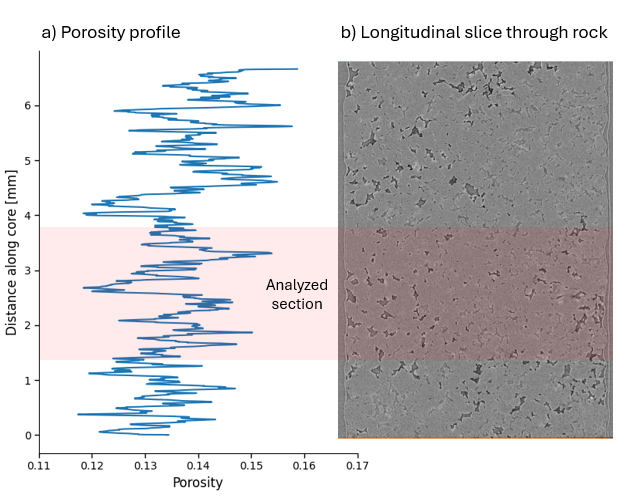}
    \caption{a) Slice-averaged porosity profile of the full imaged volume, averaged perpendicular to the flow direction. The analysed subvolume is highlighted in red, and this is the full segmented volume. b) Longitudinal slice through the rock (light grey) showing hydrogen (dark grey), brine (mid grey), scaled to align with the porosity profile and provide context for the pore-scale structure of the analysed region.}
    \label{fig:porosity}
\end{figure}

The porosity distribution across the entire imaged volume is presented in Figure \ref{fig:porosity}. The porosity ranges from 11.7\% to 15.9\%, with an average porosity of 13.6\% calculated from the segmented images. This is consistent with previously reported values for Clashach sandstone which were between 12 and 18\% \cite{jangda2024subsurface, ngwenya2003permeability, thaysen2023pore, ngwenya2000experimental}.

The 3D subvolume presented in this manuscript is smaller than the full segmented volume to improve the clarity and interpretability of the visualizations. All analyses were performed on this subvolume, while equivalent results obtained from the complete segmented volume are provided in the Supplementary Material. Comparison between the selected subvolume and the larger segmented volume (i.e., between Figures \ref{fig:time_statisticsv2} and \ref{fig:largervolume}) suggests that the selected subvolume is representative of the sample with respect to both hydrogen saturation and the number of distinct hydrogen ganglia. Thus, the selected subvolume can be considered a Representative Elementary Volume (REV), as these quantities remain comparable across subvolumes of different sizes \cite{jackson2020representative}. A full discussion of the representativeness of the selected subvolume is provided in the Supplementary Material.

The Euler characteristic ($\chi$), defined as the linear combination of the number of isolated objects ($N$), redundant loops ($L$), and cavities ($O$),

\begin{equation}
\chi = N - L + O,
\end{equation}

provides a measure of phase connectivity through its relationship with the integral of Gaussian curvature \cite{armstrong2019porous}. Unlike volumetric measurements used to calculate saturation, $\chi$ is more sensitive to segmentation methodology and inherent uncertainties \cite{spurin2024python}. It also exhibits greater variability across the subvolumes of different sizes, a reflection of its sensitivity to small changes in phase classification. 

The volume-weighted mean ganglion size was calculated to quantify the characteristic size of the hydrogen ganglia while accounting for their contribution to the total hydrogen volume. Unlike the arithmetic mean, which weights all ganglia equally regardless of size, the volume-weighted mean emphasizes larger ganglia that contain a greater proportion of the stored hydrogen. This metric is, therefore, less sensitive to the presence of numerous small disconnected clusters and provides a more representative measure of the characteristic hydrogen ganglion size. The volume-weighted mean ganglion size, $\bar{V}_w$, was calculated as

\begin{equation}
\bar{V}_w = \frac{\sum_{i=1}^{N} V_i^2}{\sum_{i=1}^{N} V_i},
\end{equation}

where $V_i$ is the volume of ganglion $i$ and $N$ is the total number of ganglia. This expression weights each ganglion by its volume, such that larger ganglia contribute proportionally more to the mean than smaller ganglia.

\section{Results and Discussion}

In this work, fast, in-situ synchrotron X-ray tomography was used to directly visualise hydrogen recovery under progressively increasing brine injection rates. By resolving pore-scale fluid redistribution during active displacement, this approach enables us to: (1) determine the onset of hydrogen mobilisation during brine invasion; (2) evaluate how fluid connectivity governs hydrogen recovery and how it evolves with flow rate and time; (3) quantify the influence of  heterogeneity on recovery; and (4) establish the relationship between hydrogen recovery and brine flow rate.

\subsection{The onset of hydrogen mobilisation during brine invasion}

The onset of dynamics, which marks the mobilisation of hydrogen and the initiation of its displacement, is critical for determining the brine injection rate required for recovery and assessing the long-term consequences of the selected flow conditions. Investigating these dynamics provides mechanistic insight into the processes governing hydrogen redistribution and mobilisation, enabling more reliable predictions of hydrogen recovery performance.

\begin{figure}[!h]
    \centering
    \includegraphics[width=1.0\textwidth]{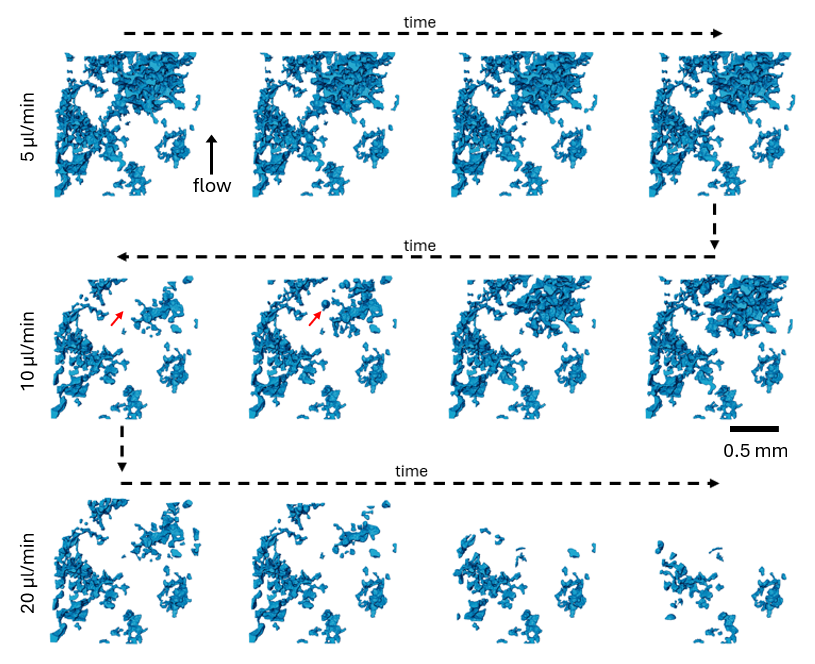}
    \caption{Hydrogen recovery over time, with the brine and rock matrix rendered transparent for visualization. Each row corresponds to a constant imposed brine flow rate. Images are shown at approximately 7, 14, 28, and 34 minutes after the start of injection, such that each row represents the same elapsed injection time.. For these experiments, the total injected brine volumes corresponded to 1.33, 2.67, and 5.33 pore volumes (PV) for the low-, intermediate-, and high-flow-rate experiments, respectively.}
    \label{fig:hydrogen_recovery_time}
\end{figure}

The 3D distributions of hydrogen during brine invasion for the different brine flow rates are shown in Figure \ref{fig:hydrogen_recovery_time}. From this, in the top row, it can be observed that little hydrogen movement occurred during the lowest brine flow rate (\SI{5}{\micro\litre\per\minute}). This behaviour is consistent with a viscous-capillary balance, where mobilisation occurs only when the imposed pressure gradient is sufficient to overcome local capillary entry-pressure barriers. When the brine flow was increased to \SI{10}{\micro\litre\per\minute}, significant displacement occurred, with constant hydrogen removal between scans (shown in the middle panel in Figure \ref{fig:hydrogen_recovery_time}, and quantified in Figure \ref{fig:time_statisticsv2}). Hydrogen ganglia that are disconnected from the inlet, and might appear trapped, are displaced by the brine. In particular, the third image acquired at \SI{10}{\micro\litre\per\minute} contains a spherical ganglion in the upper central region (see red arrow in Figure \ref{fig:hydrogen_recovery_time}) that, based on its isolated morphology, could be assumed to have formed through snap-off trapping. However, the ganglion disappears in the following scan, indicating that it remained mobile rather than being permanently trapped. This demonstrates that topological disconnection alone is insufficient to confirm trapping, as isolated ganglia may still be remobilised when local pressure gradients provide a driving force for displacement, even at low bulk capillary numbers. Similar behaviour has been observed in other fluid systems, where local capillary numbers can be substantially higher than those inferred from the average flow field, resulting in a more complex interplay of non-local and cooperative displacement events \cite{datta2014mobilization, zarikos2018velocity, rucker2015connected}. When the brine flow rate was increased to \SI{20}{\micro\litre\per\minute}, changes in ganglion size and cluster distribution were already evident between the first two images in the bottom panel of Figure \ref{fig:hydrogen_recovery_time}. Continued injection resulted in substantial hydrogen mobilisation, leading to the complete displacement of hydrogen from the upper portion of the analysed region. This indicates that there are regions of the core in which hydrogen was completely recovered, consistent with observations of complete hydrogen recovery in fractured porous media \cite{manoorkar2026microscale}.

Overall, hydrogen displacement was strongly dependent on the imposed brine flow rate. The characteristic timescales of imbibition also varied with flow rate, with hydrogen mobilisation occurring more rapidly at higher injection rates. Notably, hydrogen clusters that appeared trapped at lower flow rates were remobilised when the flow rate was subsequently increased, demonstrating that apparent trapping under one flow condition does not necessarily represent permanent immobilisation. It should be noted that the flow-rate experiments were performed sequentially on the same sample. Consequently, the fluid configuration established during one flow-rate stage influenced the initial conditions for the subsequent stage. Owing to the hysteretic nature of multiphase flow and its dependence on saturation history, the observed displacement dynamics are therefore path dependent. Complementary experiments in which each flow rate is applied to an initially hydrogen-saturated sample would be valuable for isolating the influence of flow rate from saturation history. A further consideration is that each flow-rate stage was conducted for the same duration rather than the same injected pore volume. As a result, the total injected volumes differed between experiments, which should be considered when comparing hydrogen recovery and displacement across flow rates.

Imbibition dynamics have previously been shown to persist over longer timescales than drainage dynamics for other fluid pairs \cite{ellman2025multiphase}. The onset of these dynamics, and their relationship to hydrogen connectivity within the pore space, are quantified and discussed in the following section.

\subsection{The impact of fluid connectivity on hydrogen recovery}

\begin{figure}[!h]
    \centering
    \includegraphics[width=1.0\textwidth]{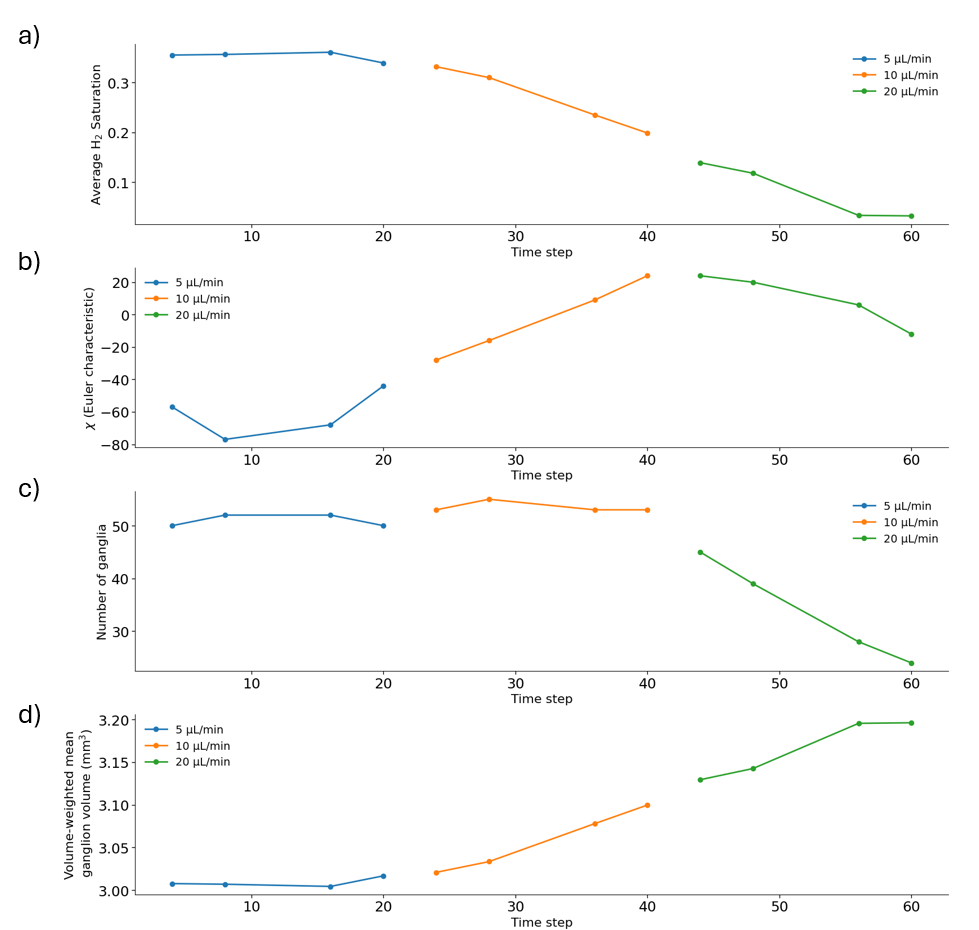}
    \caption{a) hydrogen saturation with time for different brine invasion rates, b) Euler characteristic with time for different brine invasion rates, c) number of distinct hydrogen ganglia with time for different brine invasion rates, and d): volume-weighted mean ganglion volume with time for different brine invasion rates. Each time step is uniform, and corresponds to approximately 1.7~mins. The same plot but with pore volumes injected on the x-axis is shown in Figure \ref{fig:time_stats_PV}.}
    \label{fig:time_statisticsv2}
\end{figure}

Hydrogen connectivity, and its relationship to hydrogen recovery, provides important insight into the regions of the pore space where hydrogen is preferentially mobilised and displaced. The dynamics observed in the previous section can therefore be quantified through metrics that characterise the evolution of hydrogen connectivity. The evolution of hydrogen saturation is shown qualitatively in Figure \ref{fig:hydrogen_recovery_time}, and quantitatively Figure \ref{fig:time_statisticsv2}a. As described in the previous section, the hydrogen saturation remained approximately constant during \SI{5}{\micro\litre\per\minute} brine invasion. Then there was a steady decrease during \SI{10}{\micro\litre\per\minute} brine invasion. The hydrogen saturation decreased further when the brine invasion rate was increased to \SI{20}{\micro\litre\per\minute}, with almost all the hydrogen removed from the analysed section by the end of the experiment. 

Although the hydrogen saturation was relatively stable during brine invasion at \SI{5}{\micro\litre\per\minute}, there were some fluctuations in the Euler characteristic (shown in Figure \ref{fig:time_statisticsv2}b). The Euler characteristic is a useful measure of connectivity, but is sensitive to image resolution and biases in image segmentation (see Appendix for discussion on Representative Elementary Volumes and the sensitivity of the Euler characteristic) \cite{spurin2024python}. As a result, the number of ganglia (Figure \ref{fig:time_statisticsv2}c) and the volume-weighted mean ganglion volume (Figure \ref{fig:time_statisticsv2}d) are also shown. The volume-weighted mean ganglion volume represents the average size of disconnected hydrogen ganglia while placing greater emphasis on larger clusters. Unlike a simple arithmetic mean, this metric weights each ganglion by its volume, making it more sensitive to the contribution of larger connected regions within the pore space. Physically, this metric is important because larger ganglia can span several pores and experience a larger pressure difference along their length, whereas smaller isolated ganglia are more strongly controlled by local pore-throat barriers. As a result, the metric provides insight into whether the hydrogen phase is dominated by a few large connected ganglia or by many smaller disconnected clusters during imbibition and recovery processes. Both the number of ganglia and the volume-weighted mean ganglion volume  show minor variations for the \SI{5}{\micro\litre\per\minute} brine invasion. This supports the qualitative assessment in Figure \ref{fig:hydrogen_recovery_time}.

When the brine flow rate was increased to \SI{10}{\micro\litre\per\minute}, substantial hydrogen displacement was observed qualitatively in Figure~\ref{fig:hydrogen_recovery_time} and quantitatively in Figure \ref{fig:time_statisticsv2}. Concurrently, the Euler characteristic increased from negative to positive values, indicating that the hydrogen phase became progressively more disconnected. The number of ganglia increased slightly relative to the previous flow rate, but remained comparatively stable throughout the scans acquired during \SI{10}{\micro\litre\per\minute} brine invasion. In contrast, the volume-weighted mean ganglion volume increased. When considered alongside the ganglion count, this behaviour suggests that the connected hydrogen phase initially fragmented into smaller disconnected regions, after which the smallest ganglia were preferentially mobilised and displaced from the system. At the final, and highest, brine invasion rate (\SI{20}{\micro\litre\per\minute}), hydrogen saturation decreased further; however, the Euler characteristic now decreased and became negative once again. Simultaneously, the number of ganglia declined significantly, while the volume-weighted mean ganglion volume continued to increase. This behaviour indicates that, as substantial hydrogen displacement occurred, smaller ganglia were preferentially removed from the pore space, leaving behind fewer but comparatively larger disconnected hydrogen clusters, thereby increasing the volume-weighted mean ganglion volume. This effect may be more pronounced due to the sequential increase in brine injection rate, as the sample was not resaturated between flow-rate stages. Therefore, the observed response may reflect pathway-dependent behaviour associated with saturation history, which should be considered when extrapolating these results.

These observations suggest that hydrogen recovery proceeds through two distinct stages. The first stage is characterised by the fragmentation of the initially connected hydrogen phase into smaller ganglia, leading to an increase in phase disconnectivity. The second stage involves the mobilisation and displacement of these disconnected clusters, ultimately resulting in minimal hydrogen trapping. This behaviour demonstrates that recovery is governed not only by saturation changes, but also by the evolving topology and connectivity of the hydrogen phase. The mobilisation of disconnected hydrogen clusters may have occurred at lower flow rates given a sufficiently long injection period, or the increase in brine invasion rate may have enabled the mobilisation, suggesting a rate-dependent control on hydrogen remobilisation. Longer-duration experiments are warranted to further investigate this behaviour and constrain the mechanisms governing the progressive mobilisation of disconnected hydrogen. Consequently, these findings emphasise the importance of carefully accounting for pore-scale and scale-dependent processes when upscaling laboratory observations to reservoir-scale predictions. They further highlight the need to design experiments that more closely represent the flow conditions and timescales encountered at the field scale.

\begin{figure}[!htb]
    \centering
    \includegraphics[width=1\textwidth]{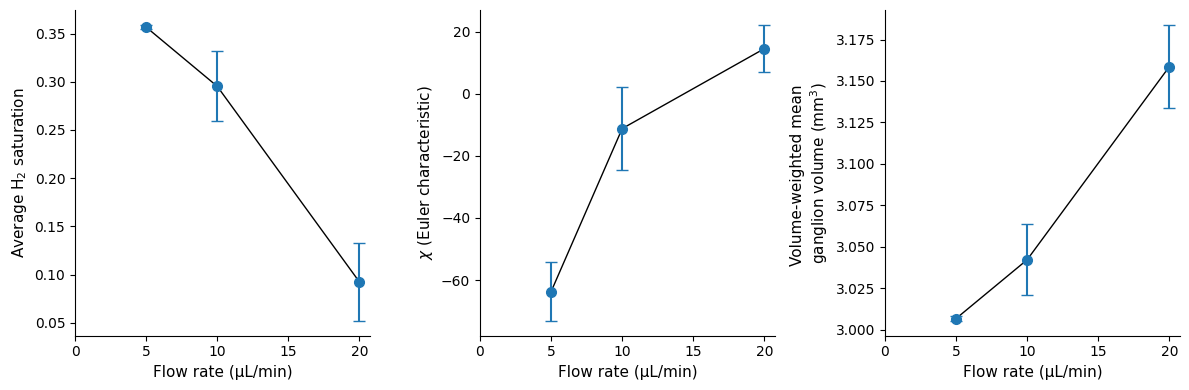}
    \caption{Average hydrogen saturation, Euler characteristic, and volume-weighted mean ganglion volume as a function of brine injection rate. Values represent the time-averaged metrics over all time steps, with error bars indicating the minimum and maximum values observed during the corresponding time series.}
    \label{fig:flowrate_analysis}
\end{figure}

The time-averaged values of the average hydrogen saturation, Euler characteristic, and volume-weighted mean ganglion volume from Figure \ref{fig:time_statisticsv2} are plotted as a function of brine injection rate in Figure \ref{fig:flowrate_analysis}. Average hydrogen saturation decreases with increasing flow rate, although the response is sub-linear: a fourfold increase in flow rate results in only approximately a $\times$1.5 reduction in saturation. The Euler characteristic increases from negative to positive values, indicating a transition from a connected to a fragmented hydrogen topology. At the same time, there is an increase in the volume-weighted mean ganglion volume reflecting the preferential mobilization and removal of smaller hydrogen ganglia. These observations support a coupled mechanism in which increasing the brine injection rate enhances hydrogen fragmentation, after which the resulting smaller ganglia are preferentially mobilized and displaced from the system. The evolution of all three metrics appears to be continuous across the range of brine injection rates investigated, with no evidence of abrupt transitions. However, as only three injection rates were investigated, further experiments spanning a broader range of flow conditions are required to fully characterise the transition in displacement dynamics and quantify their dependence on the initial hydrogen saturation.

Overall, the results demonstrate that both hydrogen connectivity and ganglion size exert a strong influence on imbibition dynamics. The evolution of the Euler characteristic, ganglion number, and volume-weighted mean ganglion volume indicates that imbibition progresses through multiple stages. The initial stage is dominated by the fragmentation of the connected hydrogen phase into smaller disconnected ganglia, increasing phase disconnectivity, whereas the subsequent stage is characterised by the mobilisation and displacement of these smaller ganglia from the pore space. This two-stage displacement behaviour has also been observed in other fluid systems using micromodel experiments \cite{datta2014mobilization}.

\subsection{The impact of heterogeneity on hydrogen recovery}

\begin{figure}[!h]
    \centering
    \includegraphics[width=1.0\textwidth]{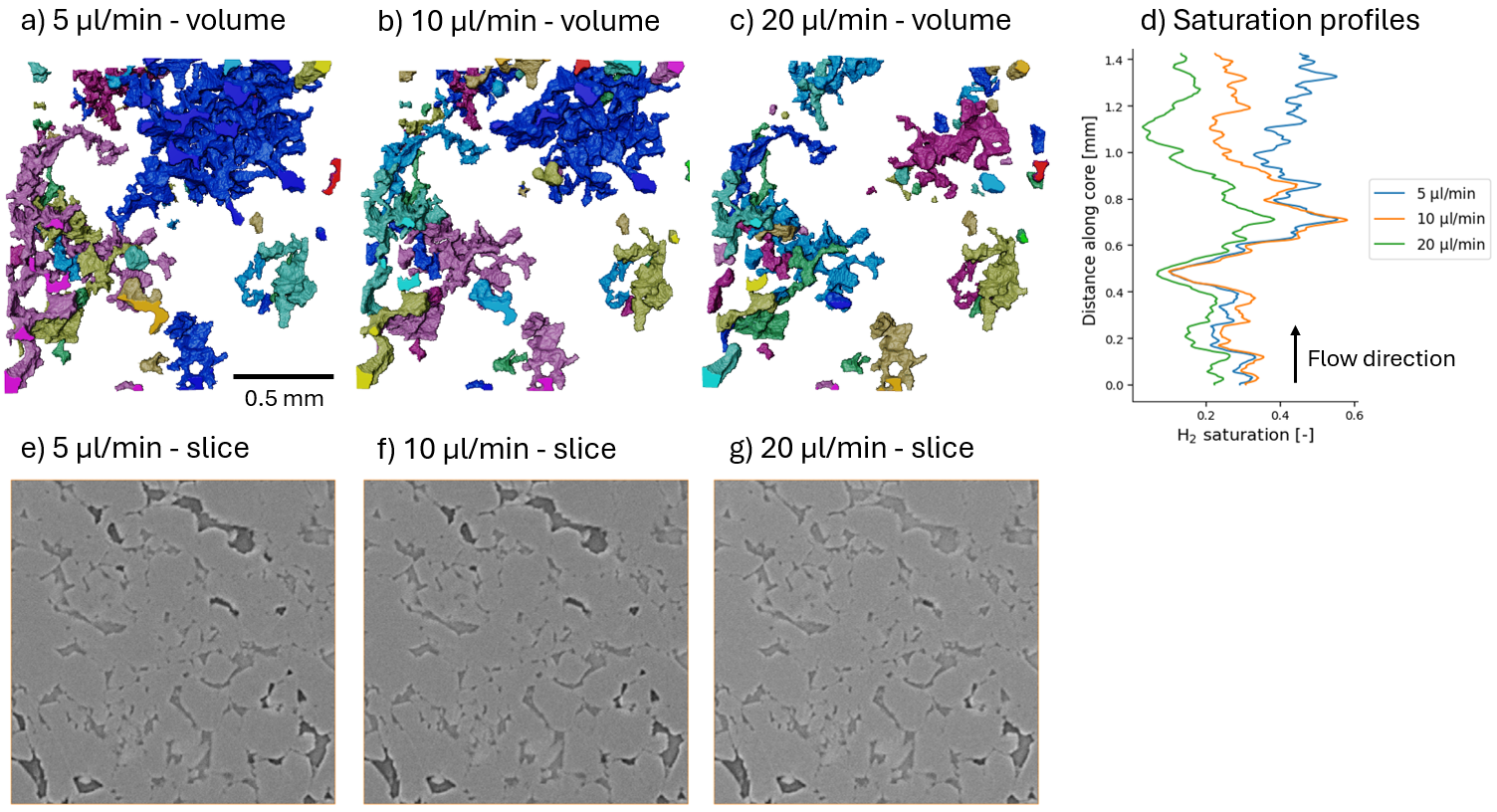}
    \caption{Hydrogen distribution for different brine flow rates in the analysed subvolume for the first time step presented in Figure \ref{fig:time_statisticsv2}: a) \SI{5}{\micro\litre\per\minute}, b) \SI{10}{\micro\litre\per\minute} and c) \SI{20}{\micro\litre\per\minute}. Only the hydrogen is shown, with the rock and brine transparent. Each distinct ganglion is given a different colour to highlight connectivity qualitatively. The saturation along the subvolume is shown in d). A slice through the rock, perpendicular to flow is shown for each flow rate in e-g), with hydrogen as the darkest phase present.}
    \label{fig:3D_different_flow_rates}
\end{figure}

Although the slice-averaged porosity profile is relatively uniform, ranging from 12-16\% (see Figure \ref{fig:porosity}), the spatial distribution of hydrogen becomes increasingly heterogeneous during brine invasion, as illustrated in Figure \ref{fig:3D_different_flow_rates}, with the corresponding slice-averaged hydrogen saturation presented in Figure \ref{fig:3D_different_flow_rates}d. The porosity variation along the core is limited to approximately $\pm$17\% of the mean porosity; however, hydrogen saturation varies by approximately $\pm$80\% of the mean at the two lowest flow rates and by $\pm$120\% at the highest flow rate. These differences indicate that similar bulk porosity values can correspond to distinct pore-scale arrangements and capillary properties, leading to preferential displacement pathways and heterogeneous hydrogen trapping during brine invasion. The lower viscosity of hydrogen increases the sensitivity of flow to small-scale core heterogeneities, promoting preferential flow pathways and highly heterogeneous displacement behaviour, consistent with observations reported in previous studies \cite{zhang2026pore, jangda2024subsurface}.

Significant hydrogen displacement is observed near the top of the imaged subvolume in Figure \ref{fig:3D_different_flow_rates}d, approximately 0.6~mm from the inlet along the imaged domain. This displacement occurs downstream of a region with lower initial hydrogen saturation, suggesting a potential capillary barrier that influences hydrogen redistribution during imbibition. The exceptionally large saturation gradients at the highest flow rate arise from extensive hydrogen displacement in the upper portion of the core. This suggests that flow dynamics are controlled by capillary heterogeneity not reflected in the porosity, or very minor variations in porosity lead to large variations in saturation. Because capillary entry pressure depends strongly on effective pore-throat radius, small variations in throat size and connectivity can produce large differences in brine reinvasion even where slice-averaged porosity appears relatively uniform.

As indicated in Figure \ref{fig:3D_different_flow_rates}d, brine is injected from the bottom of the core. A distinct increase in hydrogen saturation occurs at approximately 0.6~mm along the subvolume, corresponding to a large hydrogen ganglion spanning multiple pores in the upper-right region of Figure \ref{fig:3D_different_flow_rates}a. At higher brine invasion rates, hydrogen displacement is concentrated upstream of this feature. Although a region of elevated hydrogen saturation remains near this barrier at approximately 0.6~mm along the core, saturation decreases markedly further upstream. Conversely, little variation in hydrogen saturation is observed downstream across all flow-rate conditions. This behaviour indicates that hydrogen recovery is governed by localized pore-scale rearrangement and mobilisation events rather than spatially uniform displacement across the sample. 

\begin{figure}[!htbp]
    \centering
    \includegraphics[width=0.6\textwidth]{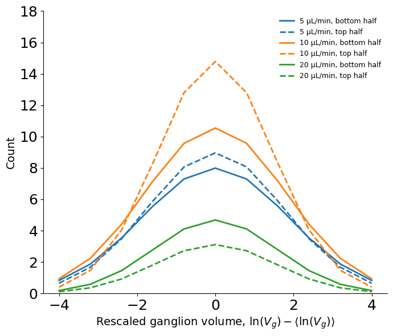}
    \caption{The ganglia distribution for the different brine injection rates. To assess the influence of the constriction, the subvolume shown in Figure \ref{fig:3D_different_flow_rates} was divided into two equal halves, approximately representing the regions above and below the constriction visible in Figure \ref{fig:3D_different_flow_rates}d. The ganglion volume as the ganglion volume varied between the different brine flow rates.}
    \label{fig:constriction_analysis}
\end{figure}

To investigate this behaviour further, the subvolume was divided into two halves (top and bottom), and the ganglion size distributions within each region are presented in Figure \ref{fig:constriction_analysis} \cite{anastasiou2024steady}. The number of ganglia increased between the \SI{5}{\micro\litre\per\minute} and \SI{10}{\micro\litre\per\minute} brine invasions, before decreasing during the \SI{20}{\micro\litre\per\minute} brine invasion. At \SI{5}{\micro\litre\per\minute}, the upper half contained a slightly greater number of ganglia than the lower half. This difference became more pronounced at \SI{10}{\micro\litre\per\minute}, indicating preferential ganglion fragmentation within the upper region. These newly formed ganglia were subsequently mobilised and redistributed, resulting in a reduction in the number of ganglia above the constriction during the \SI{20}{\micro\litre\per\minute} brine invasion. Overall, the displacement process is strongly spatially heterogeneous, with fragmentation of the large ganglion located in the upper-right region contributing to an increased ganglion population predominantly above the constriction.

%Ganglia size distribution upstream and downstream of the barrier. 

\subsection{The relationship between hydrogen recovery and brine flow rate}

\begin{figure}[!htbp]
    \centering
    \includegraphics[width=1\textwidth]{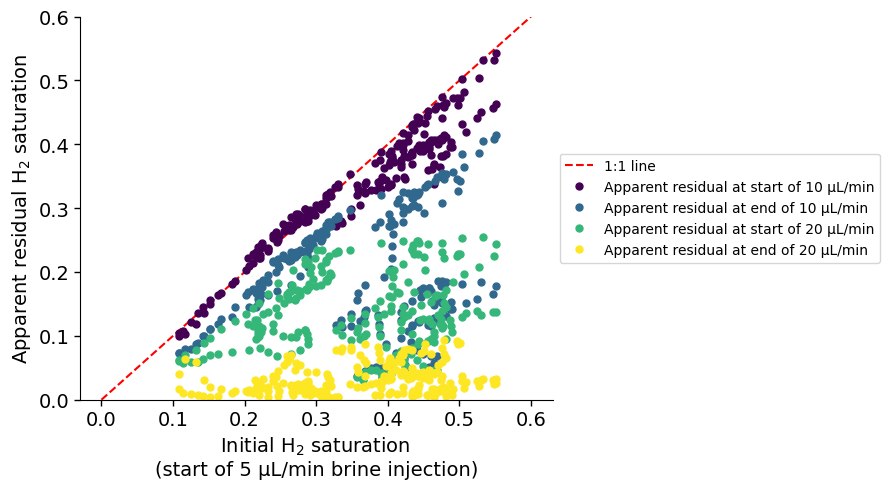}
    \caption{Hydrogen recovery during different stages of the experiment. The initial saturation is the hydrogen saturation at the start of scanning i.e. the start of the \SI{5}{\micro\litre\per\minute} brine flow. The apparent residual saturation is the saturation at the end of the different brine flow rates. While not the true residual, the term is used for consistency with convention. Each point represents the slice-averaged hydrogen saturation in the direction of flow.}
    \label{fig:trapping_eff}
\end{figure}

To investigate the influence of brine injection rate on hydrogen recovery, the hydrogen saturation at the start of the lowest brine injection rate (defined here as the initial H$_2$ saturation) is compared with the final hydrogen saturation after each brine injection rate (defined as the apparent residual H$_2$ saturation), as shown in Figure \ref{fig:trapping_eff}. This apparent residual saturation is not the true residual saturation, but is used to fit in line with traditional trapping efficiency curves and highlight regions where hydrogen is favourably being displaced during recovery. As can be observed in Figure \ref{fig:trapping_eff}, almost all the hydrogen at the start of the \SI{5}{\micro\litre\per\minute} brine invasion scans was present at the start of the \SI{10}{\micro\litre\per\minute} brine invasion. However, by the end of the \SI{20}{\micro\litre\per\minute} brine invasion almost all the hydrogen has been displaced from the system. It can also be observed in Figure \ref{fig:trapping_eff}, that regions with higher initial hydrogen saturation actually decrease in trapping efficiency first. This is shown by the significant overlap in the points at high initial hydrogen saturations, whereas the different points (colour coded for different flow rates) have less overlap for lower initial hydrogen saturations. The recovery pathway is, therefore, hysteretic: brine reinvasion does not simply reverse the preceding hydrogen drainage pathway, but is controlled by the connectivity and capillary barriers established during the initial gas invasion. Consequently, the measured recovery behaviour depends on the sequential increase in brine flow rate adopted in this study. If the sample were fully resaturated between successive flow-rate experiments, the amount and spatial distribution of trapped hydrogen would likely differ because each displacement would begin from a different initial saturation configuration. This should be considered when interpreting the effect of brine flow rate on hydrogen recovery.

Overall, almost all the hydrogen that was present prior to brine invasion was displaced when the maximum brine flow rate matched the maximum hydrogen flow rate used to saturate the sample, which is encouraging for hydrogen storage. Although the rapid loss of larger ganglia is most consistent with advective mobilisation, the gradual reduction of small disconnected clusters may also include dissolution or diffusive mass transfer, which cannot be fully separated from the present images alone. We observed that hydrogen trapping at the end of brine invasion was low, even in areas that had higher hydrogen saturation at the beginning of brine invasion. While lower recovery has been reported by other researchers, these observations were attributed to larger-scale flow bypassing, which was not observed these experiments \cite{boon2022experimental, bahrami2026pore}. This discrepancy may be scale-dependent and should therefore be considered upscaling these results to the field-scale.

\section{Conclusions and implications for hydrogen storage}

In this work, we investigated the dynamics of hydrogen recovery during brine imbibition over a range of brine injection rates. The results demonstrate that hydrogen recovery is governed by complex pore-scale displacement processes that evolve over time and are strongly influenced by both flow conditions and rock heterogeneity. We found that:

\begin{enumerate}
    \item Disconnected hydrogen clusters were remobilized during subsequent brine injection, demonstrating that disconnected ganglia are not necessarily permanently trapped. Significant hydrogen redistribution and mobilization occurred during imbibition, indicating that hydrogen recovery may be underestimated if the remobilization of initially trapped clusters is neglected. Notably, hydrogen mobilization was observed even at low bulk capillary numbers ($Ca < 10^{-6}$).
    \item Sample heterogeneity was found to strongly influence hydrogen trapping behaviour during the early stages of brine imbibition. Certain regions exhibited preferential displacement and localized trapping behaviour. Variations in saturation were not correlated with porosity heterogeneity, implying capillary pressure heterogeneity controls dynamics regardless of the flow rate. However, the influence of heterogeneity became less pronounced during the later stages of the experiment, where the residual hydrogen saturation converged towards a more spatially uniform distribution. While hydrogen distribution is heterogeneous after drainage, recovery becomes more spatially uniform during brine imbibition.
    \item The experiments demonstrate that substantial hydrogen recovery can be achieved when the brine injection rate is comparable to the original hydrogen injection rate. These findings suggest that hydrogen recovery depends strongly on the magnitude and timing of brine reinvasion relative to the preceding hydrogen invasion pathway, and that higher imbibition rates can promote the mobilisation of previously disconnected hydrogen clusters. While lower recovery has been reported by other researchers, these observations were attributed to larger-scale flow bypassing, which was not observed these experiments \cite{boon2022experimental, bahrami2026pore}. This discrepancy may be scale-dependent and should, therefore, be considered when extrapolating these results to field-scale applications.
    \item Significant time-dependent dynamics were observed throughout the imbibition process. The characteristic timescales associated with hydrogen redistribution and recovery were relatively long and evolved as a function of both time and injection rate. Future experiments should be designed with sufficient duration to capture time-dependent pore-scale dynamics relevant to field-scale recovery. Field-scale extrapolation should therefore be based on pressure gradients, capillary conditions, pore structure and saturation history, rather than direct comparison with laboratory flow rates alone.
\end{enumerate}

Significant hydrogen displacement was observed at higher brine injection rates, even when brine injection was initially performed at a low flow rate which resulted in limited hydrogen mobilization. As the brine flow rate was subsequently increased, previously immobile hydrogen ganglia were remobilized and recovered. These observations demonstrate that hydrogen remaining after low-rate injection should not necessarily be regarded as irrecoverably trapped, and that substantial additional recovery can be achieved through progressively increasing brine flow rates. This indicates that ramping the injection rate does not inherently reduce hydrogen recovery efficiency. Rather, the results emphasise the role of extended imbibition timescales in controlling displacement and recovery dynamics, which should be accounted for in the design of subsurface hydrogen storage strategies.

Future work should investigate whether stepwise brine injection strategies lead to greater hydrogen recovery compared with directly applying the maximum injection rate. Since the displacement dynamics continue to evolve over extended timescales, sufficient equilibration time must be allowed before assessing residual hydrogen saturation. Short-duration experiments may, therefore, underestimate the ultimate recoverable hydrogen and overestimate the amount of permanently trapped hydrogen.

\section{Open Research}
The segmentation workflow is available on Github: \url{https://github.com/cspurin/image_processing}. The data is available on XX. 

\section{Acknowledgements}
Thaysen, Butler, Hassanpouryouzband and Edlmann gratefully acknowledge the funding support from the Engineering and Physical Science Research Council (EPSRC) HyStorPor project [grant number EP/S027815/1]. This work was enabled and carried out with the support of Diamond Light Source, Beamline I12-JEEP 101006632 (proposal mg26730-1). Edlmann gratefully acknowledges support from the HyDRA Project, which has received funding from the European Union’s Horizon Europe Research and Innovation Programme under Grant Agreement No. 101192337, and from the Marie Skłodowska-Curie Doctoral Network “SHINE – Safe underground Hydrogen storage IN porous subsurface reservoirs” under Grant Agreement No. 101073271. Additional support was provided by UK Research and Innovation (UKRI) through the Engineering and Physical Sciences Research Council (EPSRC) under Grant Agreement EP/X026957/1.

\section{Appendix}

\subsection{Representative Elementary Volume}
\begin{figure}[!htbp]
    \centering
    \includegraphics[width=1\textwidth]{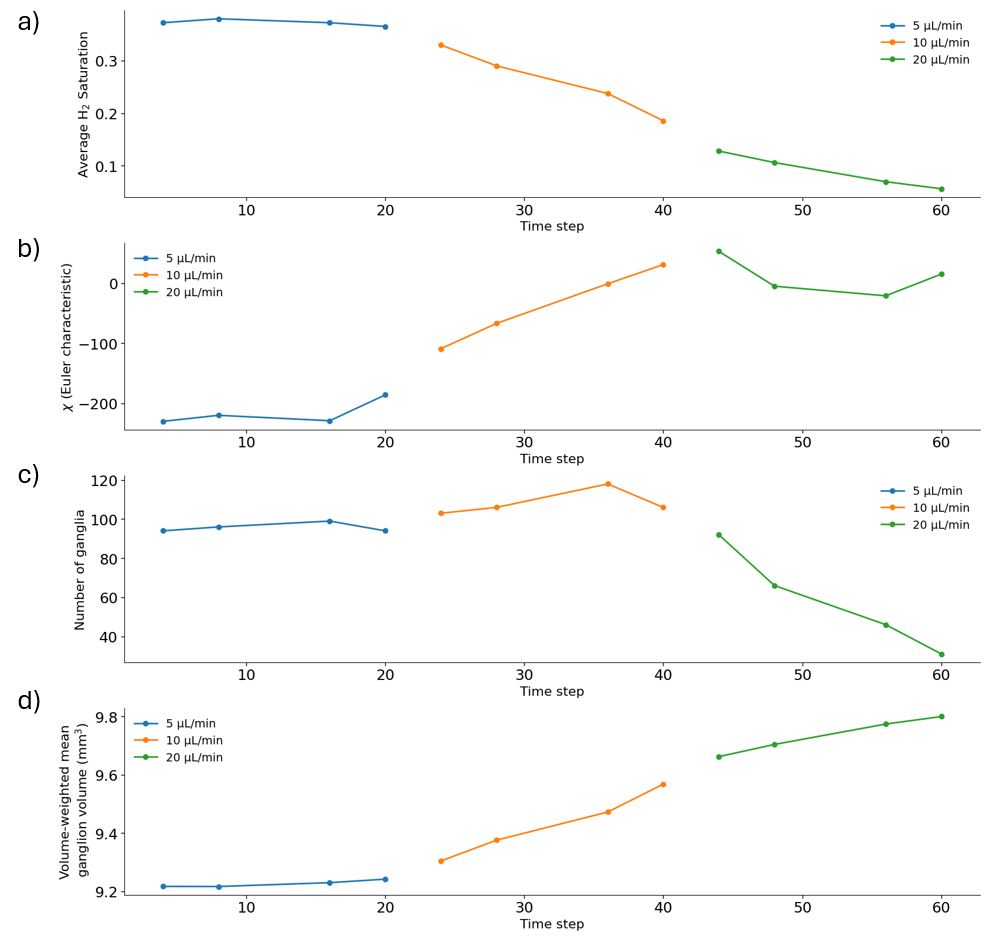}
    \caption{a) hydrogen saturation with time for the different brine invasion rates, b) Euler characteristic with time for the different brine invasion rates, c) number of distinct hydrogen ganglia with time for the different brine invasion rates, and d) volume-weighted mean ganglion volume with time for different the brine invasion rates. This is the same metrics as Figure \ref{fig:time_statisticsv2} but performed on the full segmented volume. }
    \label{fig:largervolume}
\end{figure}

The statistics for a larger volume are shown in Figure \ref{fig:largervolume}. The evolution of saturation, number of ganglia and volume-weighted mean ganglion volume have the same trends. The evolution of Euler characteristic was similar, with it remaining relatively constant and negative during \SI{5}{\micro\litre\per\minute} brine invasion, and then increasing during \SI{10}{\micro\litre\per\minute} brine invasion to indicate the hydrogen becoming more disconnected. Conversely, the Euler characteristic is flatter during the highest brine invasion rate. These observations indicate that the imaged subvolume is representative of the larger system, and that the smaller subvolume used for visualisation purposes remains representative as well. Of the metrics analysed, the Euler characteristic is the most sensitive to variations in phase connectivity and topology, while the remaining measures are comparatively more robust to segmentation choices and associated image-processing uncertainties.

\subsection{Evolution as a function of injected pore volumes}

\begin{figure}[!htbp]
    \centering
    \includegraphics[width=1\textwidth]{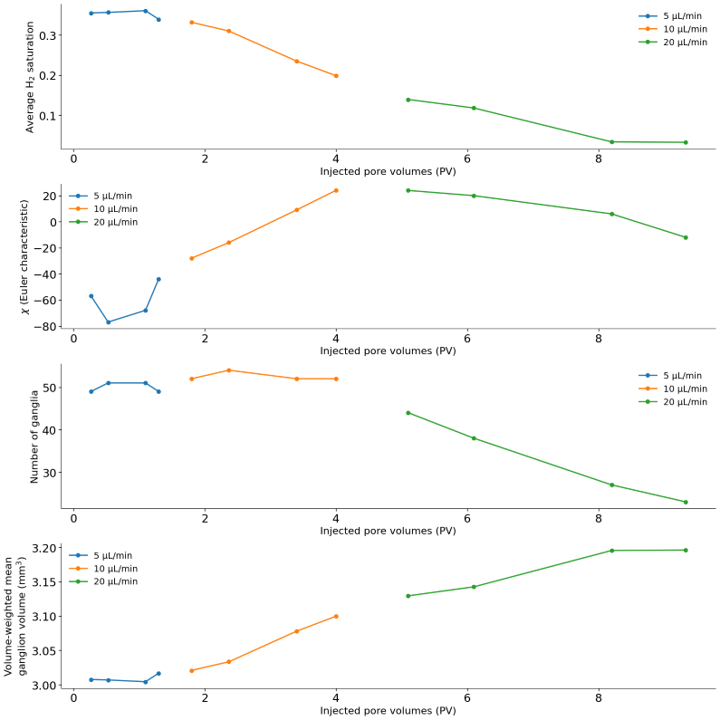}
    \caption{a) hydrogen saturation with time for the different brine invasion rates, b) Euler characteristic with time for the different brine invasion rates, c) number of distinct hydrogen ganglia with time for the different brine invasion rates, and d) volume-weighted mean ganglion volume with time for different the brine invasion rates. This is the same metrics as Figure \ref{fig:time_statisticsv2} but with the injected pore volumes on the x-axis instead of timestep. }
    \label{fig:time_stats_PV}
\end{figure}

Figure \ref{fig:time_stats_PV} presents the same data as Figure \ref{fig:time_statisticsv2}, but with cumulative injected pore volumes (PV) replacing timestep on the x-axis. Expressing the results as a function of injected PV highlights that the lower-flow-rate experiments required a greater number of pore volumes to be injected over the same experimental duration than the higher-flow-rate experiments. This difference should be considered when comparing displacement dynamics and hydrogen recovery across flow rates.

%% References and Citations
\newpage
\bibliography{refs_used_only}

\end{document}